%% file: main.tex
\documentclass[longbibliography,twocolumn,prl,amsmath,amssym,amsthm,amsfonts,superscriptaddress]{revtex4-1}
\usepackage[colorlinks=true,urlcolor=blue,citecolor=blue,linkcolor=blue]{hyperref}
\usepackage{graphicx}
\usepackage{hyperref}
\usepackage{bm}
\usepackage{newfloat}
\usepackage{makecell}
\usepackage{algorithm}
\usepackage{algcompatible}

\begin{document}
\title{Learning entropy production via neural networks}
\author{Dong-Kyum Kim}
\thanks{These authors equally contributed to this work.}
\affiliation{Department of Physics, Korea Advanced Institute of Science and Technology, Daejeon 34141, Korea}
\author{Youngkyoung Bae}
\thanks{These authors equally contributed to this work.}
\affiliation{Department of Physics, Korea Advanced Institute of Science and Technology, Daejeon 34141, Korea}
\author{Sangyun Lee}
\affiliation{Department of Physics, Korea Advanced Institute of Science and Technology, Daejeon 34141, Korea}
\author{Hawoong Jeong}
\email{hjeong@kaist.edu}
\affiliation{Department of Physics, Korea Advanced Institute of Science and Technology, Daejeon 34141, Korea}
\affiliation{Center for Complex Systems, Korea Advanced Institute of Science and Technology, Daejeon 34141, Korea}

\begin{abstract}
This Letter presents a neural estimator for entropy production, or NEEP, that estimates entropy production (EP) from trajectories of relevant variables without detailed information on the system dynamics. For steady state, we rigorously prove that the estimator, which can be built up from different choices of deep neural networks, provides stochastic EP by optimizing the objective function proposed here. We verify the NEEP with the stochastic processes of the bead-spring and discrete flashing ratchet models, and also demonstrate that our method is applicable to high-dimensional data and can provide coarse-grained EP for Markov systems with unobservable states.
\end{abstract}

\maketitle

Nonequilibrium states are ubiquitously observed from colloidal particles to biological systems~\cite{martin2001comparison, ben2011effective, weber2012nonthermal, battle2015intracellular, battle2016broken, gladrow2016broken}. Injection of energy, lack of relaxation time, or broken detailed balance are ordinary sources of nonequilibrium, and in general, such systems are in contact with a heat bath such as a fluid. Thus, to describe the behavior of a nonequilibrium system, it is necessary to investigate the energetics of the system; however, experimentally, heat flow is difficult to measure directly~\cite{harada2005equality, toyabe2007experimental,  lander2012noninvasive, gnesotto2018broken}. In this case, measuring the entropy production (EP) can be one remedy to estimate heat flow in a nonequilibrium system~\cite{sekimoto2010stochastic, seifert2012stochastic, gnesotto2018broken}.

Many techniques have been developed to accurately measure EP, such as approaches calculating probability currents and density~\cite{lander2012noninvasive,  Li2019quantifying}. These methods require detailed information from a governing equation though, so to address this issue, a few methods to estimate the EP rate without such detailed information have been proposed, including the plug-in method~\cite{wang2005divergence, Roldan2010Estimating, roldan2012entropy}, the compression-based estimator~\cite{ziv1993measure, Roldan2010Estimating, roldan2012entropy, Avinery2019Universal, martiniani2019quantifying}, and the thermodynamic uncertainty relation (TUR) based estimator~\cite{barato2015thermodynamic, manikandan2019inferring, Li2019quantifying, van2020entropy, otsubo2020estimating}. The plug-in and compression-based methods estimate the EP rate through the Kullback--Leibler divergence, but they are only applicable for discrete state variables. And while the TUR-based approach has recently been adopted in frameworks for the exact estimation of EP rates and distributions in short time limits~\cite{manikandan2019inferring, van2020entropy, otsubo2020estimating}, estimating stochastic EP remains an unsolved issue for continuous state variables.

Various fields in physics have been employing machine learning (ML) to solve a wide range of non-trivial problems such as identifying relevant variables~\cite{mehta2014exact, koch2018mutual, li2019neural}, identifying phase transitions~\cite{carrasquilla2017machine, van2017learning, venderley2018machine, liu2018discriminative, beach2018machine, zhang2018machine}, quantum many-body problems~\cite{carleo2017solving, deng2017quantum, chng2017machine, choo2018symmetries, torlai2018neural, hartmann2019neural, nagy2019variational, vicentini2019variational}, and others~\cite{carleo2019machine}. Likewise, ML has also been applied to EP rate estimation~\cite{gnesotto2020learning, otsubo2020estimating} as well as classification of the direction of time's arrow~\cite{seif2019machine}. Relatedly, in the ML community, a recent work by Rahaman et al.~\cite{Rahaman2020Learning} proposed a neural network to measure an entropy-like quantity by unsupervised learning; however, the quantity was not physically well defined, i.e. it had no scale. To the best of our knowledge, estimating EP using neural networks has yet to be explored.

\begin{figure}[!b]
\includegraphics[width=\columnwidth]{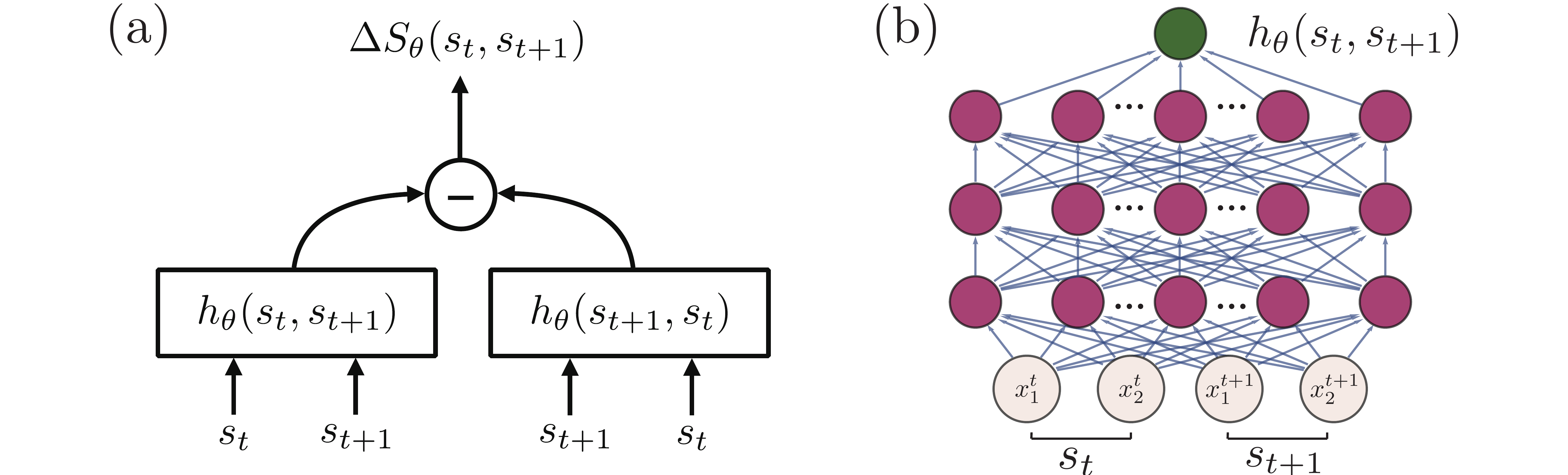}
\vskip -0.1in
\caption{(a) Architecture of the neural estimator for entropy production (NEEP). (b) Illustration of a multilayer perceptron (MLP) with three hidden layers for an $N=2$ bead-spring model where $s_t=(x^t_1, x^t_2)$.}
\label{fig:fig1}
\end{figure}

\begin{figure*}[!t]
\centering
\includegraphics[width=\textwidth]{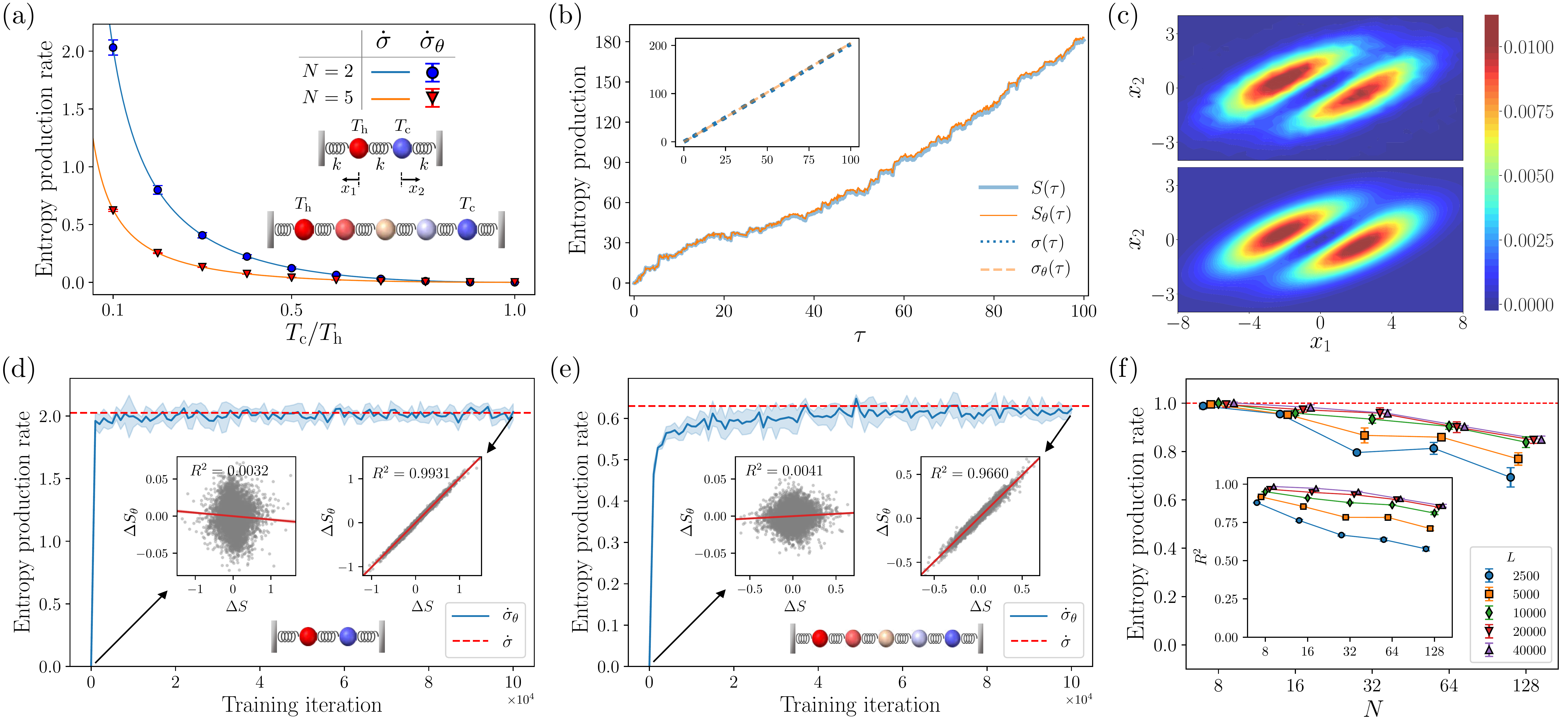}
\vskip -0.1in
\caption{(a) Entropy production (EP) rate as a function of $T_{\rm c}/T_{\rm h}$ for models with two and five beads. The solid lines (symbols) indicate the analytical EP rate $\dot{\sigma}$ (estimated EP rate $\dot{\sigma}_{\theta}$). (b) Cumulative EP over time $\tau$ along a single trajectory, which is randomly sampled from the test set. The inset shows the ensemble-averaged EP. (c) Local EP rate as a function of $x_1$ and $x_2$. The top and bottom panels show the NEEP and the analytical results, respectively. (d,e) $\dot{\sigma}_{\theta}$ with respect to training iteration for (d) two beads and (e) five beads. The left (right) inset corresponds to results before (after) training, showing scatter plots between $\Delta S$ and $\Delta S_{\theta}$ with a fitted linear regression line (solid red line). The results in (b--e) are performed at $T_{\rm c}/T_{\rm h} = 0.1$. (f) Results of NEEP for high-dimensional bead-spring models. $\dot{\sigma}_{\theta}$ as a function of $N$ for each number of steps ($L$) are plotted with five different markers, as indicated in the legend. The $ R^2 $ values of the linear regression between $\Delta S$ and $\Delta S_{\theta}$ are shown in the inset. The red dashed line denotes $\dot{\sigma}$. Error bars and shaded areas represent the standard deviation of estimations from five independently trained estimators.}
\label{fig:bead-spring}
\vskip -0.1in
\end{figure*}

In this Letter, we propose the neural estimator for entropy production (NEEP), which can estimate stochastic EP from the time-series data of relevant variables without detailed information on the dynamics of the system. For Markov chain trajectory $s_1, s_2, ..., s_L$, we build a function $h_{\theta}$ that takes two states, $s_t$ and $s_{t+1}$, where $\theta$ denotes the trainable neural network parameters. As shown in Fig.~\ref{fig:fig1}(a), the output of NEEP is defined as
\begin{align}
    \Delta S_{\theta}(s_t, s_{t+1}) \equiv h_{\theta}(s_t, s_{t+1}) - h_{\theta}(s_{t+1}, s_t). \label{eq:estimator}
\end{align}
Here, $\Delta S_{\theta}(s_t, s_{t+1})$ satisfies the antisymmetric relation $\Delta S_{\theta}(s_t, s_{t+1})= -\Delta S_{\theta}(s_{t+1}, s_t)$.
We define the objective function to be maximized as
\begin{align}
    J(\theta) &= \mathbb{E}_t \mathbb{E}_{s_t \to s_{t+1}} [\Delta S_{\theta}(s_t, s_{t+1}) - e^{-\Delta S_{\theta}(s_t, s_{t+1})}],
    \label{eq:lossftn}
\end{align}
where $\mathbb{E}_t$ denotes the expectation over $t$, which is uniformly sampled from $\{1, ..., L-1\}$, and $\mathbb{E}_{s_t \to s_{t+1}}$ is the expectation over transition $s_t \to s_{t+1}$.
If detailed balance is satisfied, then the transition $s \to s'$ and its reverse transition $s' \to s$ equally appear in the ensemble of the trajectories. In this case, the optimized $\Delta S_{\theta}$ is zero for all possible transitions, but if detailed balance is broken, then $\Delta S_{\theta}$ becomes larger due to more irreversible transitions.
In steady state, $J(\theta)$ can be written as
\begin{align}
    J[h] &=\sum_{i, j} 
        p_{i} T_{ij} \left[ (h_{ij} - h_{ji}) - e^{-(h_{ij} - h_{ji})}
    \right]
    \label{eq:NESSloss}
\end{align}
where we set $h_{ij} \equiv h(s_i, s_j)$, $p_{i}\equiv p(s_i)$ is the steady-state probability density, and 
$T_{ji}\equiv p(s_i,t+1|s_j,t)$ is a propagator. Because the neural networks tune output $h_{\alpha\beta} \equiv h(s_{\alpha},s_{\beta})$ by optimizing $\theta$, the maximum condition for Eq.~\eqref{eq:NESSloss} becomes
\begin{align}\label{eq:derivNESSloss}
0 =& \partial_{h_{\alpha\beta}} J[h]\\
  =&\sum_{i, j} 
    \left[ 
        p_{i} T_{ij} (1 + e^{-(h_{ij}-h_{ji})} )
        (\delta_{i\alpha}\delta_{j\beta} - \delta_{i\beta}\delta_{j \alpha}) 
    \right]\nonumber\\
    =& p_{\alpha} T_{\alpha\beta} (1 +  e^{-(h_{\alpha\beta} - h_{\beta\alpha})}) \nonumber - p_{\beta} T_{\beta\alpha}(1 +
    e^{-(h_{\beta\alpha} - h_{\alpha\beta})}). \nonumber
\end{align}
Then the solution for the optimization problem is 
\begin{align}
    h_{\alpha\beta} - h_{\beta\alpha} = -\ln{(p_{\beta} T_{\beta \alpha} / p_{\alpha} T_{\alpha \beta} ),}
    \label{eq:sol}
\end{align}
which is the definition of stochastic entropy production~\cite{seifert2012stochastic} when $T_{ji} = \widetilde{T}_{ij}$. Here, $\widetilde{T}_{ij}$ is the time-reversal propagator of $T_{ij}$. This proof supports the ability of our NEEP to learn appropriate EP. We maximize Eq.~\eqref{eq:lossftn} via the stochastic gradient ascent method that is widely used in deep learning literature~\cite{lecun2015deep, goodfellow2016deep}. See Supplemental Material (SM)~\cite{SM} for the training and evaluation details.

To validate our approach, we estimate the EP of two widely studied nonequilibrium systems: the bead-spring model for continuous state variables~\cite{battle2016broken, mura2018nonequilibrium, Li2019quantifying, gnesotto2020learning} and the discrete flashing ratchet model for discrete state variables~\cite{ajdari1992ratchet, Roldan2010Estimating, roldan2012entropy}. To attempt more challenging problems, we additionally apply NEEP to high-dimensional continuous models and a hidden Markov model.

In the bead-spring model, $N$ beads are coupled to the nearest beads or boundary walls by springs and contacted with thermal heat baths at different temperatures, as described in Fig.~\ref{fig:bead-spring}(a). For displacements $x_1, x_2, ..., x_N$, the dynamics of $N$-beads is governed by an overdamped Langevin equation
\begin{align}
    \dot{x}_i(\tau) =& A_{ij}x_{j}(\tau) + \sqrt{2 k_{\rm B} T_{i} / \gamma} \xi_{i}(\tau),
\end{align}
where $A_{ij} = (-2\delta_{i,j} + \delta_{i, j+1} + \delta_{i+1, j})k /\gamma$. Here, $k$ is a spring constant, $\gamma$ is the Stokes friction coefficient, and the temperature $T_i$ of each heat bath linearly varies from $T_{\rm h}$ to $T_{\rm c}$. $\xi_i$ is an independent Gaussian white noise satisfying $\mathbb{E}[\xi_{i}(\tau)\xi_{j}(\tau')] = \delta_{ij}\delta(\tau-\tau')$ where $\mathbb{E}$ denotes the ensemble average. We set all the parameters to be dimensionless and $k_B = k = \gamma = 1$. The linearly varying temperature induces a thermodynamic force that drives the system to a nonequilibrium state.

To attempt EP estimation in a system with continuous variables, we firstly consider $N=2$ and $N=5$ bead-spring models. Here, $ \dot{\sigma} $ is the analytical value of the ensemble-averaged EP rate~\cite{SM}; Fig.~\ref{fig:bead-spring}(a) plots $\dot{\sigma}$ for $N=2$~($5$) with a blue (orange) solid line. As illustrated in Fig.~\ref{fig:fig1}(b), we employ a 3-hidden-layer multilayer perceptron (MLP) for $h_{\theta}$. See SM~\cite{SM} for the configuration and robustness of the architecture. For training and test sets, we numerically sampled $10^3$ positional trajectories in steady state for each model. Each trajectory was sampled with time step $\Delta \tau = 10^{-2}$~\footnote{The time interval between the current and next states can affect the estimation of NEEP. For large enough time intervals, the correlation between the two states weakens and NEEP may provide imprecise estimations.}, and the total number of steps $L$ is $10^4$. We present the training results at $T_{\rm c}/T_{\rm h} = 0.1$ in Fig.~\ref{fig:bead-spring}(b--e). Note that all reported results in Fig.~\ref{fig:bead-spring} are from the test set. We also demonstrate the estimation ability of NEEP with various $L$~\cite{SM}.

For the $N=2$ case, as shown in Fig.~\ref{fig:bead-spring}(b), it is observed that our estimator provides accurate values not only for the ensemble average but also for a single trajectory over $\tau$. Here, $S(\tau) \equiv \sum_{i=0}^{\tau/\Delta \tau} \Delta S(s_{i}, s_{i+1})$ and $\sigma (\tau) \equiv \mathbb{E}[S(\tau)]$ where $\Delta S$ is the analytic stochastic EP per $\Delta \tau$. Figure~\ref{fig:bead-spring}(c) shows that the local EP rate over the displacement space $(x_1, x_2)$ calculated by NEEP (top panel) is the same as the analytical solution (bottom panel). The local EP rate from NEEP at $(x_1, x_2)$ is measured by averaging the EP rate produced when a particle passes through the point $(x_1, x_2)$.

To check the training process, we plot the estimated values of $\dot{\sigma}_\theta$ over training iteration in Fig.~\ref{fig:bead-spring}(d). The dashed red line indicates $\dot{\sigma}$. Insets in Fig.~\ref{fig:bead-spring}(d) are scatter plots between $\Delta S_\theta$ and $\Delta S$ in a randomly sampled single trajectory. As can be seen in the left inset, there is no correlation between $\Delta S_\theta$ and $\Delta S$ before training. But after training (right inset), $\Delta S_\theta$ is well-fitted to $\Delta S$ (coefficient of determination $R^2 = 0.9931$).

We apply the same process to the $N=5$ bead-spring model, where estimating $\Delta S$ and $\dot{\sigma}$ using the thermodynamic force is difficult due to the curse of dimensionality~\cite{Li2019quantifying}. The result shows that $\Delta S_\theta$ is again well-fitted to $\Delta S$ with $R^2 = 0.9660$ (see Fig.~\ref{fig:bead-spring}(e)). We also train our estimator at $T_{\rm c}$ in the range of 1--10 with $T_{\rm h} = 10$, as indicated in Fig.~\ref{fig:bead-spring}(a), and verify that NEEP provides the exact EP rate with small errors. Notably, these results are from the test set, implying that NEEP can be generalized to estimate EP even for unseen data.

\begin{figure}[!t]
\centering
\includegraphics[width=\columnwidth]{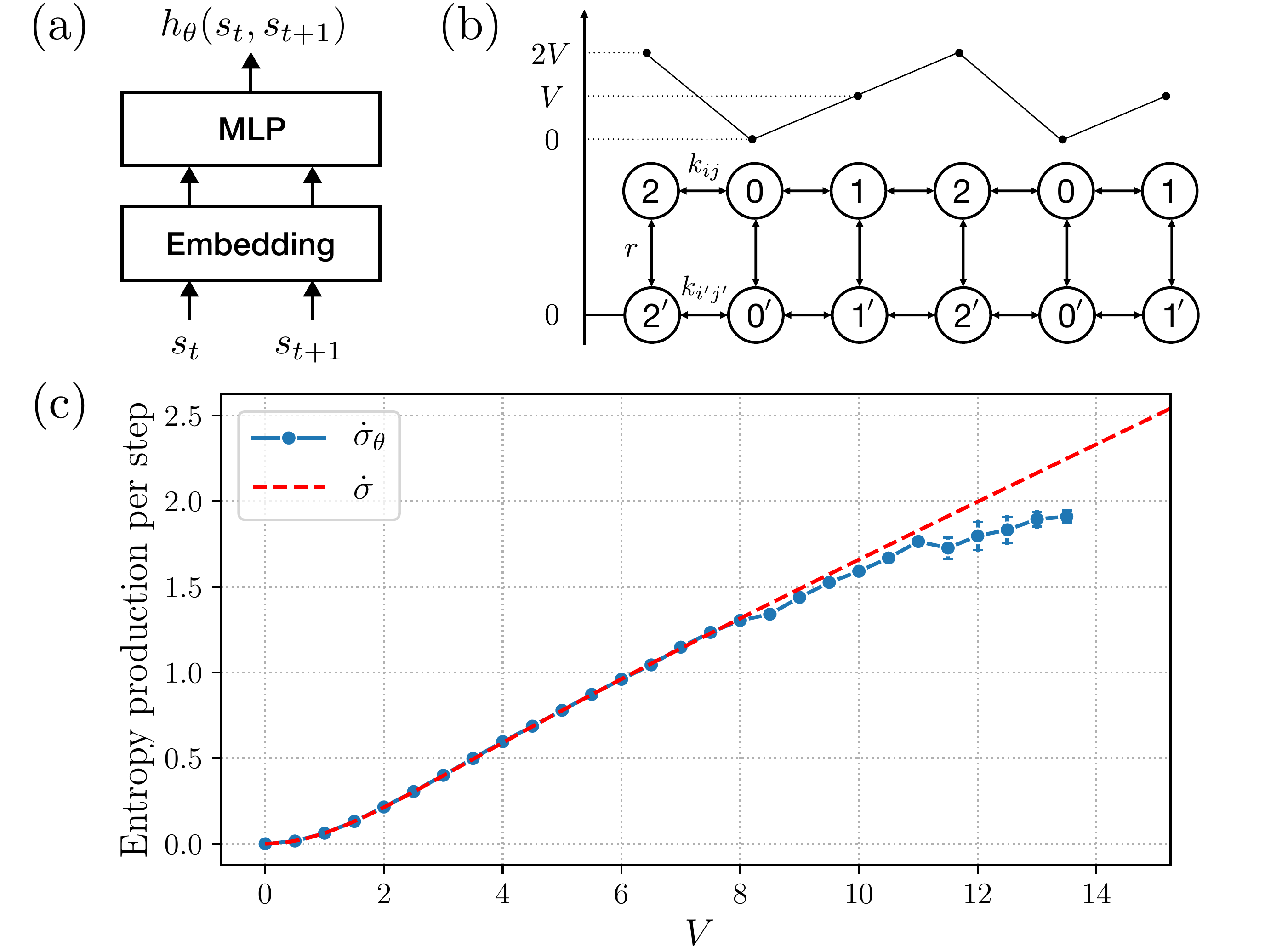}
\vskip -0.1in
\caption{(a) NEEP architecture for discrete state Markov chains. (b) Schematic of a discrete flashing ratchet model. (c) Entropy production per step as a function of potential $V$. Error bars represent the standard deviation of $\dot{\sigma}_{\theta}$ from five independently trained estimators.}
\label{fig:dfr}
\vskip -0.1in
\end{figure}

Estimating EP in high-dimensional Langevin systems has not been explored because of the curse of dimensionality~\cite{van2020entropy}. While a recent work~\cite{otsubo2020estimating} has made estimations of EP rates up to $N=15$ using TUR, here, we apply NEEP to bead-spring models with $N=8$, $16$, $32$, $64$, and $128$. For each $N$, we set $T_{\rm h}=10$ and $T_{\rm c}$ to a value where $\dot{\sigma}=1$~\cite{SM}. By increasing the training data points ($10^3L$), we can see that $\dot{\sigma}_{\theta}$ for each $N$ approaches 1 in Fig.~\ref{fig:bead-spring}(f). Although EP rate estimation errors of over 10\% are seen for $N=64$ and $128$, the $R^2$ values support that NEEP was able to learn the stochastic EP with appreciable correlations (see the inset in Fig.~\ref{fig:bead-spring}(f)). Note that, with an increasing number of beads, the architecture of NEEP does not change except for the number of input nodes ($2N$), which means that our neural estimator's computation time and the number of parameters are linearly proportional to $N$. Based on these points, we show that NEEP can efficiently mitigate the curse of dimensionality through a neural network.

Next, we demonstrate our method on the discrete flashing ratchet model~\cite{ajdari1992ratchet}, which consists of a particle moving in a one-dimensional periodic lattice. The particle is in contact with a heat bath at temperature $T$ and drifts in a periodic asymmetric sawtooth potential (see Fig.~\ref{fig:dfr}(b)). For brevity, we set $k_{B}=T=1$. In this model, the particle state has two variables, $x$ and $\eta$, where $x \in \{0, 1, 2\}$ is the position and $\eta \in \{\text{ON}, \text{OFF}\}$ is the on/off potential; the state is indicated as $i \equiv (i, \text{ON})$ and $i' \equiv (i, \text{OFF})$. Transition rates between each state $s \in \{0,1,2, 0',1',2'\}$ are defined as $k_{ij}=e^{(V_j-V_i)/2}$ and $k_{i'j'}=1$ for $i \neq j$ where $V_i$ is the potential at $i$ that switches on and off at rate $r=1$, i.e. $k_{ii'}=k_{i'i}=r$. As in a previous work~\cite{Roldan2010Estimating}, we generate a series of states and remove the information of the times when transitions occur; in this case, the analytic EP per step is given as $\dot{\sigma} = \sum_{\alpha, \beta} p(\alpha, \beta)(V_{\alpha} - V_{\beta})$.

\begin{figure}[!t]
\centering
\includegraphics[width=\columnwidth]{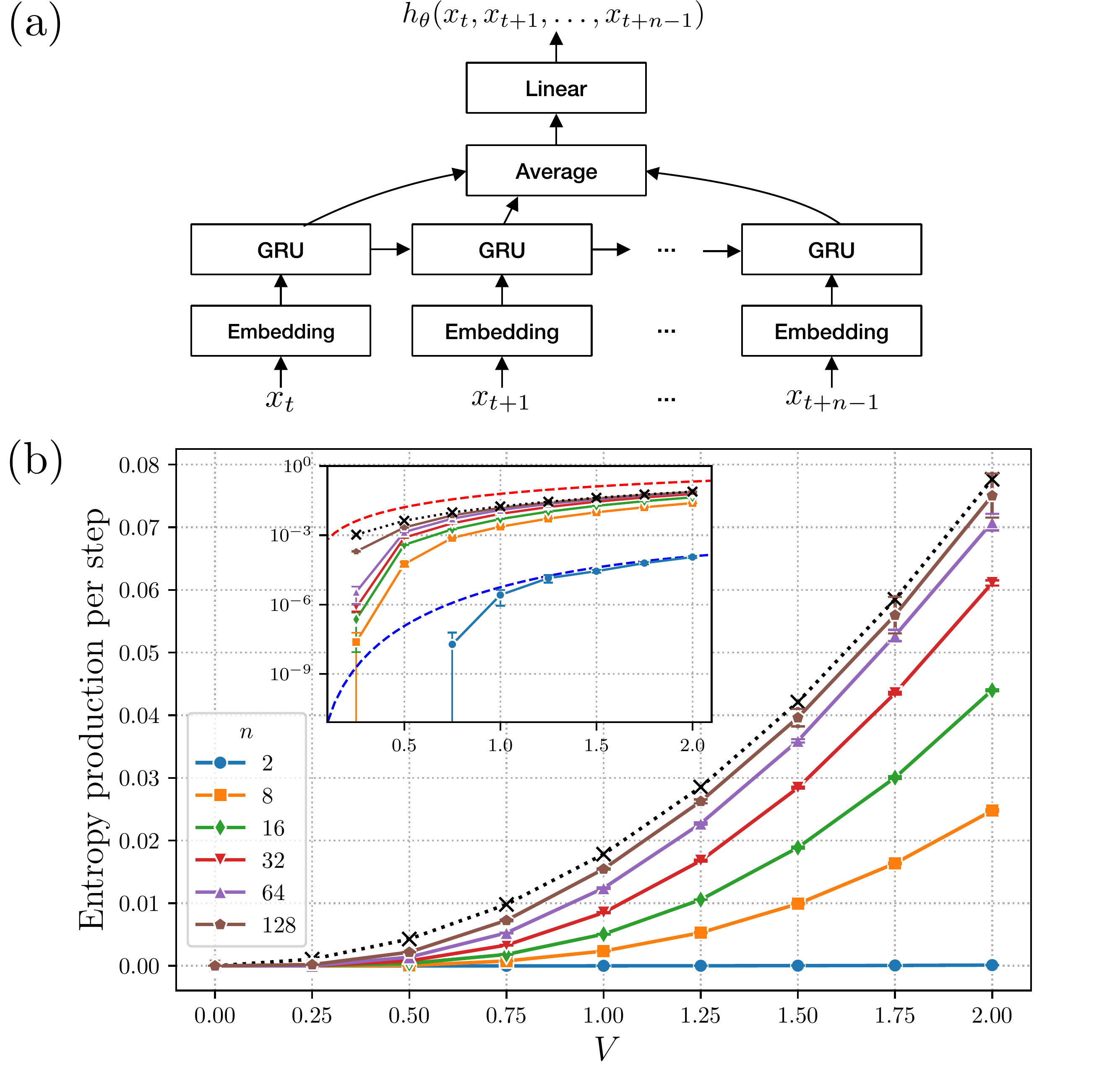}
\vskip -0.1in
\caption{(a) An RNN version of NEEP (RNEEP) for a hidden Markov model. (b) Results of RNEEP for a partial information problem. The estimations $\dot{\Sigma}^n_{\theta}$ as a function of potential $V$ for each sequence length $n$ are plotted with six different markers as shown in the legend. The black x's are the semianalytical values of $\dot{\Sigma}^{\infty}$. The inset shows a plot of the y-axis in log scale. The red (blue) dashed line denotes the analytic value of $\dot{\sigma}$ ($\dot{\Sigma}^2$). See Fig. S6 in SM~\cite{SM} for a comparison with $\dot{\sigma}$ in linear scale. Error bars represent the standard deviation of estimations from five independently trained estimators.}
\label{fig:dfr-partial}
\vskip -0.1in
\end{figure}

We construct the NEEP as shown in Fig.~\ref{fig:dfr}(a) using an embedding layer that transforms a discrete state into a trainable continuous vector called an embedding vector. After the transformation, we feed the two embedding vectors of states $s_t$ and $s_{t+1}$ to the MLP~\cite{SM}. From a set of different potential values, we sampled two single trajectories with $ L=10^6 $ steps for each potential $V$ ($0 \le V \le 15$); one trajectory is used for training and the other for testing. For the training data, we build five NEEPs, randomly initialized with five different random seeds for each potential. Figure~\ref{fig:dfr}(c) shows that $\dot\sigma$ is within the error bar of the NEEP estimations of EP per step $\dot\sigma_{\theta}$ where $V \le 8$. For $V$ in the range 8--14, the overfitting~\cite{goodfellow2016deep} problem occurs due to a lack of transitions from low to high potential, which leads to an underestimation of $\dot{\sigma}_{\theta}$ (Fig.~\ref{fig:dfr}(c)). See SM~\cite{SM} for a more detailed discussion on how we address the overfitting issue. For $14 \le V$, the probability to detect the $0 \to 2$ transition is below $0.5$ in our simulation with $L=10^6$. In this case, $\dot{\sigma}_{\theta}$ diverges because of no observation of the $0 \to 2$ transition (see Fig. S5 in SM~\cite{SM}).

So far, Markovian systems with completely observable states have been tested; however, full state information cannot often be accessed, with only some coarse-grained variables typically available. In such cases, the EP of a coarse-grained trajectory, called coarse-grained EP, is measurable~\cite{gomez2008lower, esposito2012stochastic, polettini2017effective, dabelow2019irreversibility}. To test for coarse-grained EP estimation, we assume that the on/off information $\eta$ is now inaccessible~\cite{Roldan2010Estimating, roldan2012entropy}. To address this problem, we build $h_{\theta}$ with a recurrent neural network (RNN), a popular network to consider memory effects in time-series data. We employ a gated recurrent unit (GRU)~\cite{cho-etal-2014-learning} for the RNN. As shown in Fig.~\ref{fig:dfr-partial}(a), the RNN version of NEEP (RNEEP) takes input as a series of states with a sequence length of $n$, and the outputs of the GRU are averaged over the sequence and then fed to a single layer feed-forward neural network, which is the last layer. Now, the RNEEP output is defined as $\Delta S_{\theta}(\bm{x}^n_t) \equiv h_{\theta}(\bm{x}^n_t) - h_{\theta}(\widetilde{\bm{x}}^n_t)$, and the objective function is defined as
\begin{align}
    J(\theta) &= \mathbb{E}_t \mathbb{E}_{(\bm{x}^n_t,\bm{\eta}^n_t)} [\Delta S_{\theta}(\bm{x}^n_t) - e^{-\Delta S_{\theta}(\bm{x}^n_t)}],
    \label{eq:rneeplossftn}
\end{align}
where
\begin{align}
    \bm{x}^n_t = (x_t, x_{t+1}, ..., x_{t+n-1}),~\bm{\eta}^n_t = (\eta_t, \eta_{t+1}, ..., \eta_{t+n-1}). \nonumber
\end{align}
Here, $ \widetilde{\bm{x}}^n_t$ is the time-reversed trajectory of $\bm{x}^n_t$.
In steady state, the solution for this optimization problem is the stochastic coarse-grained EP along the trajectory $\bm{x}^n$ (see SM~\cite{SM} for the proof):
\begin{align}
    \Delta S_{\theta}(\bm{x}^n) = - \ln{\frac{\sum_{\widetilde{\bm{\eta}}^n} p(\widetilde{\bm{x}}^n, \widetilde{\bm{\eta}}^n)}{\sum_{\bm{\eta}^n} p(\bm{x}^n, \bm{\eta}^n)}} = -\ln{\frac{p(\widetilde{\bm{x}}^n)}{p(\bm{x}^n)}}.
    \label{eq:coarse-grained-sol}
\end{align}
Here, the ensemble-averaged coarse-grained EP of trajectory $\bm{x}^n$ per step is denoted as $\dot{\Sigma}^n \equiv \mathbb{E}\left[-\frac{1}{n-1}\ln{\frac{p(\widetilde{\bm{x}}^n)}{p(\bm{x}^n)}}\right]$. In general, $\dot{\Sigma}^n $ provides a lower bound on the actual EP per step $\dot{\sigma}$~\cite{Roldan2010Estimating, roldan2012entropy}.

For $0 \le V \le 2$, we train the RNEEP with six different sequence lengths, $n=2,8,16,32,64$, and $128$, for maximizing Eq.~\eqref{eq:rneeplossftn} using the position trajectory with $L = 5 \times 10^7$. As can be seen in Fig.~\ref{fig:dfr-partial}(b), with increasing sequence length $n$, the estimation of RNEEP ($\dot{\Sigma}^n_{\theta}$) approaches the semianalytical value of the coarse-grained EP per step for $n \to \infty$ ($\dot{\Sigma}^{\infty}$)~\cite{roldan2012entropy}. We can verify that $\dot{\Sigma}^2_{\theta}$ is well-fitted to the analytic value of $\dot{\Sigma}^{2}$, but it remains difficult to estimate for $V \le 1$ (see the inset in Fig.~\ref{fig:dfr-partial}(b)), because the number of transitions between any two positions, e.g. $x \to y$ or $y \to x$, appears almost equally in the trajectory. While directly estimating Eq.~\eqref{eq:coarse-grained-sol} by counting the frequency of $\bm{x}^n$ is not possible for $n \ge 16$ due to the curse of dimensionality, the RNEEP can resolve this issue and enable us to estimate the coarse-grained EP up to $n=128$ (see Figs. S7--8 in SM~\cite{SM}).

In previous approaches~\cite{gnesotto2018broken}, estimation of the probability distribution and the probability current was essential to quantify how far the system is out of equilibrium. As NEEP does not require such estimation or detailed information of the system, we expect our estimator to be applicable to various fields such as active matter, biological systems, information machines, electronic devices, and others. This approach will be particularly useful to investigate the stochastic energetics and spatiotemporal patterns of dissipated energy in various systems. We further expect our method to be applicable to the understanding of complex nonequilibrium systems, e.g. soft biological assemblies~\cite{mura2018nonequilibrium} or molecular motors with hidden internal states~\cite{martinez2019inferring}. As a future work, modifying our NEEP method to estimate EP in more general nonequilibrium systems like time-dependent states will be intriguing.

The code for NEEP, implemented in \texttt{PyTorch}~\cite{pytorch2019}, is available in Ref.~\cite{kim2020repo}.

\begin{acknowledgments}
This study was supported by the Basic Science Research Program through the National Research Foundation of Korea (NRF) (KR) [NRF-2017R1A2B3006930].
\end{acknowledgments}

\input{main.bbl}
\input{supplemetal_material}

\end{document}

%% file: supplemetal_material.tex
\pagebreak
\onecolumngrid
\newpage
\begin{center}
\textbf{\large Supplemental material: Learning entropy production via neural networks}
\end{center}

\setcounter{equation}{0}
\setcounter{figure}{0}
\setcounter{table}{0}
\setcounter{page}{1}

\renewcommand{\theequation}{S\arabic{equation}}
\renewcommand{\thefigure}{S\arabic{figure}}

\newcommand{\todo}[1]{\textbf{\textcolor{red}{#1}}}

\newenvironment{psmallmatrix}
  {\left(\begin{smallmatrix}}
  {\end{smallmatrix}\right)}

\section{Training setup and algorithm}
\label{sec:experiment}
We employ ReLU~\cite{nair2010rectified} as the activation function for the neural estimator for entropy production (NEEP). We train the estimators with the Adam~\cite{kingma2014adam} optimizer using the following hyper-parameters: learning rate $10^{-4}$, weight decay $5\times10^{-5}$, and batch size 4096. We use these hyper-parameters for all processes unless noted. See Algorithm~\ref{alg:training} for our training procedure. All runs were conducted on a single NVIDIA TITAN V GPU.

\begin{algorithm}[H]
\begin{algorithmic}[1]
\REQUIRE Estimator $\Delta S_{\theta}$, optimizer, training set $\mathcal{D}_{\rm train}=\{(s^i_1, s^i_2, ..., s^i_L)\}_{i=1,..., M}$ where $L$ is the trajectory length and $M$ is the number of trajectories
\LOOP
    \STATE Compute 
    \begin{align}\label{eqs:batch_J}
        \hat{J}(\theta) = \frac{1}{|\mathcal{B}|}\sum_{(i, t) \in \mathcal{B}} \left[\Delta S_{\theta}(s^i_t, s^i_{t+1}) - e^{-\Delta S_{\theta}(s^i_t, s^i_{t+1})}\right]
    \end{align}
    where $\mathcal{B}$ is a randomly sampled subset of $\{1,..., M\} \times \{1,...,L-1\}$ where $\times$ is a Cartesian product. The total number of the sampled data points $|\mathcal{B}|$ is equal to the batch size.
    \STATE Compute gradients $\nabla_\theta \hat{J}(\theta)$.
    \STATE Update parameters $\theta$ with the optimizer.
\ENDLOOP
\end{algorithmic}
\caption{Training the NEEP}
\label{alg:training}
\end{algorithm}

\section{Architecture robustness and Evaluation metric}
\label{sec:evaluation}

\begin{figure}[!h]
\centering
\includegraphics[width=\textwidth]{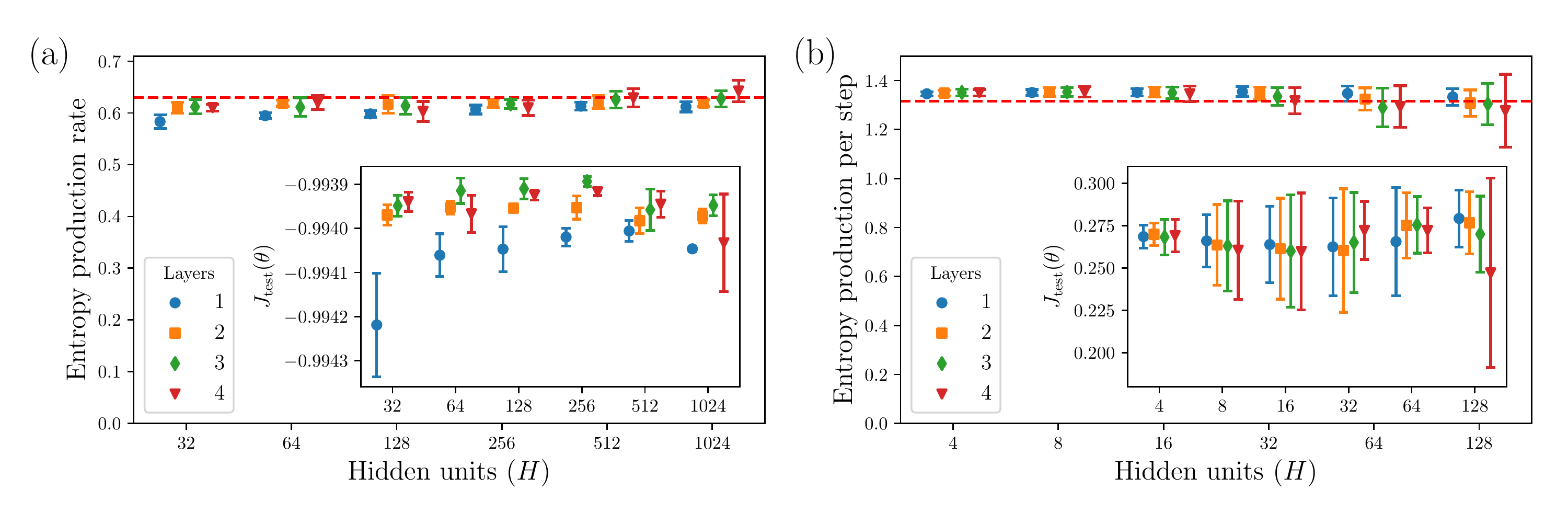}
\vskip -0.1in
\caption{Estimations on EP rate (per step) $\dot{\sigma}_{\theta}$ with multiple architecture configurations. Training and test set data are generated at (a) $T_{\rm c}/T_{\rm h}=0.1$ with $M=10^3$ and $L=10^4$ for five-bead models and at (b) $V=8$ with $M=1$ and $L=10^6$ for discrete flashing ratchet models. The insets shows the $J_{\rm test}(\theta)$ values after training. The red dashed line denotes the analytic EP rate (EP per step) $\dot{\sigma}$. Error bars represent the standard deviation of the evaluated values from five independently trained estimators.}
\vskip -0.1in
\label{figS:evaluation}
\end{figure}

The purpose of training is to get the maximum value of $J$ in general, not only for training set $\mathcal{D}_{\rm train}$, and therefore we evaluate $J_{\rm test}(\theta)$ with test set $\mathcal{D}_{\rm test}$ after training where
\begin{align}\label{eqs:test_J}
    J_{\rm test}(\theta) \equiv \frac{1}{M(L-1)} \sum_{(s^i_t, s^i_{t+1}) \in \mathcal{D}_{\rm test}} \left[\Delta S_{\theta}(s^i_t, s^i_{t+1}) - e^{-\Delta S_{\theta}(s^i_t, s^i_{t+1})}\right].
\end{align}
For the $N=5$ bead-spring model, we train NEEP built with various MLP configurations over $10^5$ training iterations: numbers of hidden layers of 1, 2, 3, and 4 and numbers of hidden units ($H$) of 32, 64, 128, 256, 512, and 1024. See Table~\ref{table:MLP} for the architecture details. After training, we evaluate the EP rate estimation $\dot{\sigma}_{\theta}$ and $J_{\rm test}(\theta)$ (Eq.~\eqref{eqs:test_J}) for each MLP configuration. Figure~\ref{figS:evaluation}(a) shows that EP rate estimations are robust to different model configurations and within a 10\% error analytic EP rate (red dashed line). The inset in Fig.~\ref{figS:evaluation}(a) shows the $J_{\rm test}(\theta)$ for each MLP configuration. As can be seen, the MLP configurations with $J_{\rm test}(\theta) < -0.9940$ either underestimate (single-layer MLPs) or overestimate (4-hidden-layer MLP with 1024 hidden units) as compared to other configurations. This supports that $J_{\rm test}(\theta)$ can evaluate whether the NEEP has learned EP well. Based on these observations, we choose the 3-hidden-layer MLP architecture with $H=256$, which shows the highest $J_{\rm test}(\theta)$ value, for $h_{\theta}$ in the paper.

For the discrete flashing ratchet model, we train NEEP built with various configurations (see Table~\ref{table:embeddingMLP}) over $5 \times 10^4$ training iterations: numbers of hidden layers of 1, 2, 3, and 4 and dimensions of the embedding vector ($H$) of 4, 8, 16, 32, 64, and 128. As with the previous results for the bead-spring model, the estimations of EP per step are robust to the configurations except for overparameterized models (see Fig.~\ref{figS:evaluation}(b)). As shown in Fig~\ref{figS:evaluation}(b) inset, we choose the single-hidden-layer MLP with $H=128$ for $h_{\theta}$ in the paper.

The large variances in $J_{\rm test}(\theta)$ and $\dot{\sigma}_{\theta}$ for overparameterized models (see the inset in Fig.~\ref{figS:evaluation}(b)) are addressed in Sec.~\ref{sec:overfitting}.

\begin{table}[H]
\centering
\begin{tabular}{l|l|l}
\hline
\multicolumn{3}{c}{
\centering
\textbf{3-hidden-layer MLP NEEP}} \\

\hline
  Layer & \makecell{Output dim} & Activation function\\
\hline
Input $(s_{t}, s_{t+1})$ & $2N$ &   \\
Fully-connected & $H$ & ReLU\\
Fully-connected & $H$ & ReLU\\
Fully-connected & $H$ &  ReLU\\
Output layer & $1$ &  None\\
\hline
\end{tabular}
\caption{Three-hidden-layer MLP configuration for $N$ bead-spring model. $H$ is the number of hidden units.}
\label{table:MLP}
\end{table}

\begin{table}[H]
\centering
\begin{tabular}{l|l|l}
\hline
\multicolumn{3}{c}{
\centering
\textbf{Embedding-MLP NEEP}} \\

\hline
  \makecell{Layer} & \makecell{Output dim} & Activation function\\
\hline
Input $s$ & $\{0, 1, 2, 0', 1', 2'\}$ & \\
Embedding & $H$ &   \\
Concatenate $(s_t, s_{t+1})$ & $2H$ &   \\
Fully-connected &  $2H$ & ReLU\\
Output layer & $1$ &  None\\
\hline
\end{tabular}
\caption{MLP with embedding layer for discrete flashing ratchet model. $H$ is the size of the embedding dimension.}
\label{table:embeddingMLP}
\end{table}

\section{Analytic description of bead-spring model}
\label{sec:analytic}
The dynamics of the beads can be described by an overdamped Langevin equation given by
\begin{align}
    \dot{x}_i(\tau) =& A_{ij}x_{j}(\tau) + \sqrt{2 k_{\rm B} T_{i} / \gamma} \xi_{i}(\tau),
\end{align}
where $A_{ij} = -2k/\gamma \; \delta_{i,j} + k/\gamma \; (\delta_{i, j+1} + \delta_{i+1, j})$, $\bm{x} = (x_{1}(\tau),x_{2}(\tau),..., x_{N}(\tau))$, $\bm{\xi} = (\xi_{1}(\tau),\xi_{2}(\tau),..., \xi_{N}(\tau))$ and the temperature $T_i$ of each heat bath linearly varies from $T_{\rm h}$ to $T_{\rm c}$. These different temperatures induce a thermodynamic force which drives the system out of equilibrium. Here, $\gamma$ is the Stokes friction coefficient, and $\bm\xi$ is an independent Gaussian white noise vector satisfying $\mathbb{E}[\xi_{i}(\tau)\xi_{j}(\tau')]= \delta_{ij}\delta(\tau-\tau')$. For brevity, we set $k_{B}=1$.
To calculate the entropy production (EP) of the bead-spring model, we have to consider the Fokker--Planck equation given by
\begin{equation}\label{Sup_eq:OverFPEq}
\begin{aligned}
    \frac{\partial p(\bm{x}, \tau)}{\partial \tau} = -\nabla \cdot \bm{j}(\bm{x}, \tau),
\end{aligned}
\end{equation}
where the probability current $\bm{j}(\bm{x}, \tau)$ is defined by
\begin{equation}\label{Sup_eq:OverCurrent}
\begin{aligned}
    \bm{j} (\bm{x}, \tau) = \mathsf{A}\bm{x}p(\bm{x}, \tau)  - \mathsf{D} \nabla p(\bm{x}, \tau). 
\end{aligned}
\end{equation}
Here, $\mathsf{D}$ is the diffusion matrix defined as $\rm{diag}\{T_1/\gamma, \dotsi, T_N/\gamma \}$.
Since our system is an Ornstein--Uhlenbeck process, the steady-state probability density function is Gaussian as $p(\bm{x}) \propto \exp[-(1/2)\bm{x}^T \mathsf{C^{-1}}\bm{x}]$ with a covariance matrix $\mathsf{C}$. 
Using the Lyapunov equation $\mathsf{A}\mathsf{C} + \mathsf{C}\mathsf{A^T} = -2\mathsf{D}$, the probability density function and the probability current in steady state, $p(\bm{x})$ and $j_{ss}(\bm{x})$, can be obtained.

The EP rate along a trajectory is given by~\cite{seifert2005entropy}
\begin{equation}\label{Sup_eq:EP_def}
\begin{aligned}
    \dot{S}(\tau) &= -\frac{\partial_\tau p(\bm{x}, \tau)}{p(\bm{x}, \tau)} \Big{|}_{\bm{x}(\tau)} + \frac{\bm{j}(\bm{x}, \tau)^T \mathsf{D}^{-1}}{p(\bm{x}, \tau)}\Big{|}_{\bm{x}(\tau)} \dot{\bm{x}}.
\end{aligned}
\end{equation}
Because the first term on the right-hand side vanishes in steady-state, the ensemble averaged entropy production rate is obtained by
\begin{equation}\label{Sup_eq:EP_def2}
\begin{aligned}
    \dot{\sigma} = \int d\bm{x} \; \frac{\bm{j}_{ss}(\bm{x})^T \mathsf{D}^{-1}}{p(\bm{x})} \bm{j}_{ss}(\bm{x}) = \mathsf{Tr} \left[ \mathsf{D}^{-1}\mathsf{A}\mathsf{C}\mathsf{A^T} - \mathsf{C}^{-1}\mathsf{D} \right],
\end{aligned}
\end{equation}
where $\mathsf{Tr}[\cdot]$ is the trace operator. We used the Lyapunov equation to derive the last expression in Eq.~\eqref{Sup_eq:EP_def2}.

For $N=2$, the deterministic term $\mathsf{A} \equiv \frac{k}{\gamma}\begin{psmallmatrix} -2 & 1 \\ 1 & -2 \end{psmallmatrix}$ and the diffusion matrix $\mathsf{D} \equiv \frac{1}{\gamma}\begin{psmallmatrix} T_1 & 0 \\ 0 & T_2 \end{psmallmatrix}$. The covariance matrix can be derived by the Lyapunov equation as
\begin{equation}\label{Sup_eq:Cov_mat}
\begin{aligned}
    \mathsf{C}  = \frac{1}{12k}\begin{pmatrix} 7T_{\rm h} + T_{\rm c} & 2(T_{\rm h} + T_{\rm c}) \\ 2(T_{\rm h} + T_{\rm c}) & T_{\rm h} + 7 T_{\rm c} \end{pmatrix}.
\end{aligned}
\end{equation}
Using Eq.~\eqref{Sup_eq:Cov_mat}, $\dot{\sigma}$ for $N=2$ is obtained as
\begin{equation}\label{eq:EP_2beads}
\begin{aligned}
    \dot{\sigma} = \frac{k(T_{\rm h} - T_{\rm c})^2}{4 \gamma T_{\rm h} T_{\rm c}}.
\end{aligned}
\end{equation}
In the same way, $\dot{\sigma}$ for the $N=5$ bead-spring model can be obtained as
\begin{equation}\label{eq:EP_5beads1}
\begin{aligned}
    \dot{\sigma} = \frac{k(T_{\rm h} - T_{\rm c})^2 (111T_{\rm h}^2 + 430T_{\rm h} T_{\rm c} + 111T_{\rm c}^2)}{495 \gamma T_{\rm h} T_{\rm c} (3T_{\rm h} + T_{\rm c})(T_{\rm h} + 3T_{\rm c})}.
\end{aligned}
\end{equation}
For $N=8$, $16$, $32$, $64$, and $128$ bead-spring models, we can also calcuate $\dot{\sigma}$ using Eq.~\eqref{Sup_eq:EP_def2}.

We analytically calculated the local EP rate in Fig.~\ref{fig:bead-spring}(c) (bottom) using the integrand of Eq.~\eqref{Sup_eq:EP_def2}. Because our NEEP model estimates along the time evolution of a particle, the local EP rate from NEEP $\dot{\sigma}_\theta (\bm{x})$ is measured by the following equation:
\begin{equation}\label{eq:EP_5beads2}
\begin{aligned}
    \dot{\sigma}_\theta (\bm{x}) = \frac{1}{L\Delta \tau} \sum_{t=0}^{L} \delta_{\bm{x}, \bm{x}_{t}^m}\, \Delta S_\theta(\bm{x}_{t}, \bm{x}_{t+1}),
\end{aligned}
\end{equation}
where $\Delta \tau$ is the time step of a trajectory $\bm{x}_1, ..., \bm{x}_L$ and $\bm{x}_{i}^m = (\bm{x}_t + \bm{x}_{t+1})/2$.

\section{Overfitting}
\label{sec:overfitting}

\begin{algorithm}[H]
\begin{algorithmic}[1]
\REQUIRE Estimator $\Delta S_{\theta}$, optimizer, training set $\mathcal{D}_{\rm train}$, test set $\mathcal{D}_{\rm test}$.
\REQUIRE Best model parameter $\theta^{\star}$, best value $J_{\rm best}$.

\STATE Initialize $J_{\rm best} \gets J_{\rm test}(\theta)$, $\theta^{\star} \gets \theta$.
\LOOP
    \STATE Compute $\hat{J}(\theta)$ (Eq.~\eqref{eqs:batch_J}) from $\mathcal{D}_{\rm train}$.
    \STATE Compute gradients $\nabla_\theta \hat{J}(\theta)$.
    \STATE Update parameters $\theta$ with the optimizer.
    \STATE Compute $J_{\rm test}(\theta)$ (Eq.~\eqref{eqs:test_J}) from $\mathcal{D}_{\rm test}$.
    \IF{$J_{\rm test}(\theta) > J_{\rm best}$}
        \STATE $J_{\rm best} \gets J_{\rm test}(\theta)$
        \STATE $\theta^{\star} \gets \theta$
    \ENDIF
\ENDLOOP
\end{algorithmic}
\caption{Monitoring $J_{\rm test}(\theta)$ during training}
\label{alg:overfitting}
\end{algorithm}

\begin{figure}[!h]
\centering
\includegraphics[width=\textwidth]{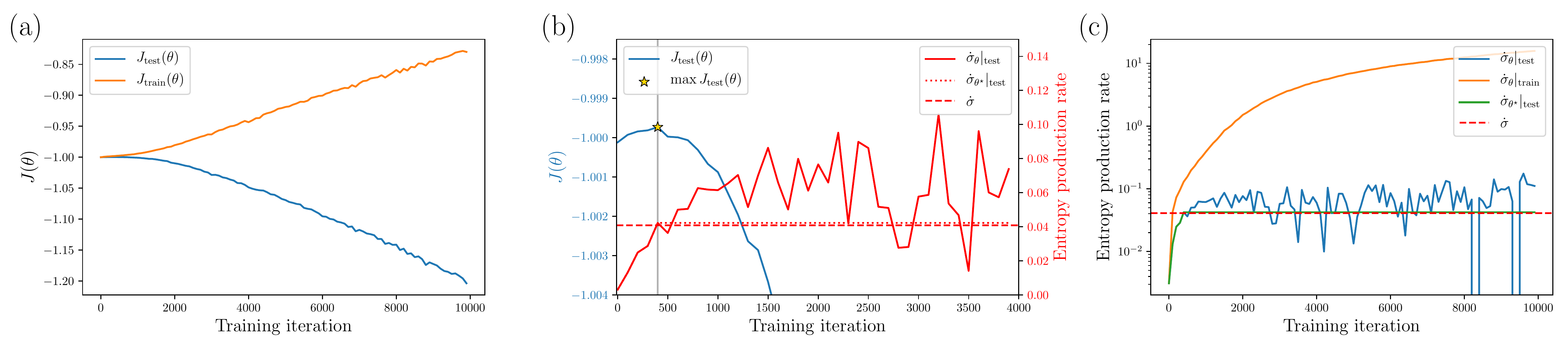}
\vskip -0.1in
\caption{(a) $J_{\rm test}(\theta)$ (blue) and $J_{\rm train}(\theta)$ (orange) with respect to training iteration. (b) $J_{\rm test}(\theta)$ and $\dot{\sigma}_{\theta}|_{\rm test}$. The yellow star shows the maximum value of $J_{\rm test}(\theta)$ during the whole training process. The red solid line denotes the EP rate estimation $\dot{\sigma}_{\theta}|_{\rm test}$. (c) The EP rate estimation with respect to training iteration. The red dashed line denotes the analytic EP rate $\dot{\sigma}$.}
\vskip -0.1in
\label{figS:overfitting}
\end{figure}

In this section, we show an example of the overfitting phenomenon that occurs when we train the NEEP for the $N=5$ bead-spring trajectory with $M=10^3$ and $L=200$ at $T_{\rm c}/T_{\rm h}=0.5$. Figure~\ref{figS:overfitting}(a) shows that the gap between $J_{\rm train}(\theta)$ and $J_{\rm test}(\theta)$ keeps increasing. This phenomenon is called overfitting~\cite{goodfellow2016deep}. To address this problem, we monitor the $J_{\rm test}(\theta)$ value during the training process. See Algorithm~\ref{alg:overfitting} for details. Steps 6--10 in Algorithm~\ref{alg:overfitting} are computationally impractical, so we only run these steps every 100 iterations. As can be seen in Fig.~\ref{figS:overfitting}(b), $J_{\rm test}(\theta)$ has a maximum value at a training iteration of 400, which is marked with a yellow star. The EP rate estimation at the maximum value $\dot{\sigma}_{\theta^{\star}}|_{\rm test}$ is represented with a red dotted line in Fig.~\ref{figS:overfitting}(b).
Figure~\ref{figS:overfitting}(c) shows that EP estimation using training set $\dot{\sigma}_{\theta}|_{\rm train}$ (orange line) keeps increasing, and EP estimation using test set $\dot{\sigma}_{\theta}|_{\rm test}$ (blue line) does not converge to a certain value; however, the $\dot{\sigma}_{\theta^{\star}}|_{\rm test}$ (green line) remains unchanged after the training iteration of 400 and is close to the analytic EP rate $\dot{\sigma}$ (red dashed line).

\begin{figure}[!h]
\centering
\includegraphics[width=\textwidth]{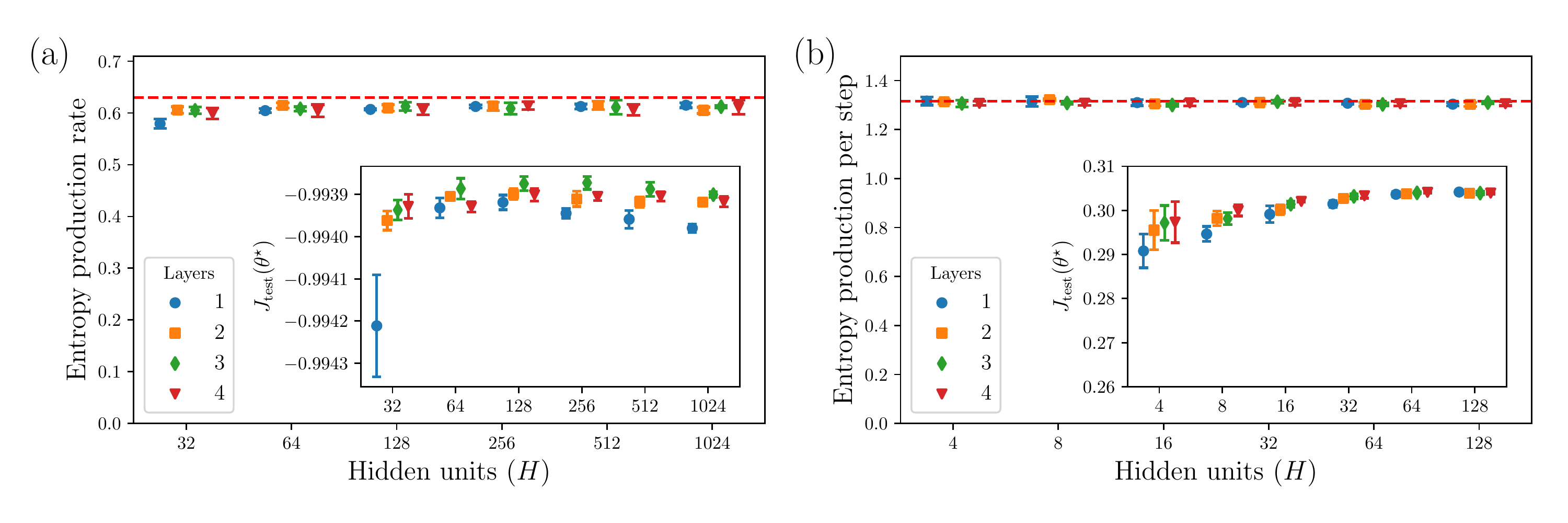}
\vskip -0.1in
\caption{Estimations on EP rate (per step) $\dot{\sigma}_{\theta^{\star}}|_{\rm test}$ with multiple architecture configurations. Training and test set data are generated at (a) $T_{\rm c}/T_{\rm h}=0.1$ with $M=10^3$ and $L=10^4$ for five-bead models and at (b) $V=8$ with $M=1$ and $L=10^6$ for discrete flashing ratchet models. The insets show the $J_{\rm test}(\theta^{\star})$ values. The red dashed lines denote the analytic EP rate (EP per step) $\dot{\sigma}$. Error bars represent the standard deviation of the evaluated values from five independently trained estimators.}
\vskip -0.1in
\label{figS:robustness}
\end{figure}

In Sec.~\ref{sec:evaluation}, there are large variances in $J_{\rm test}(\theta)$ and $\dot{\sigma}_{\theta}|_{\rm test}$ for overparameterized models. Here, we employ Algorithm~\ref{alg:overfitting} for the same runs as in Fig.~\ref{figS:evaluation}. Figure~\ref{figS:robustness} shows that the variances of $\dot{\sigma}_{\theta^{\star}}|_{\rm test}$ and $J_{\rm test}(\theta^{\star})$ are significantly smaller than before (see Fig.~\ref{figS:evaluation}). This result supports that Algorithm~\ref{alg:overfitting} can ensure robustness even for overparameterized neural network architectures.

\begin{figure}[!h]
\centering
\includegraphics[width=0.8\textwidth]{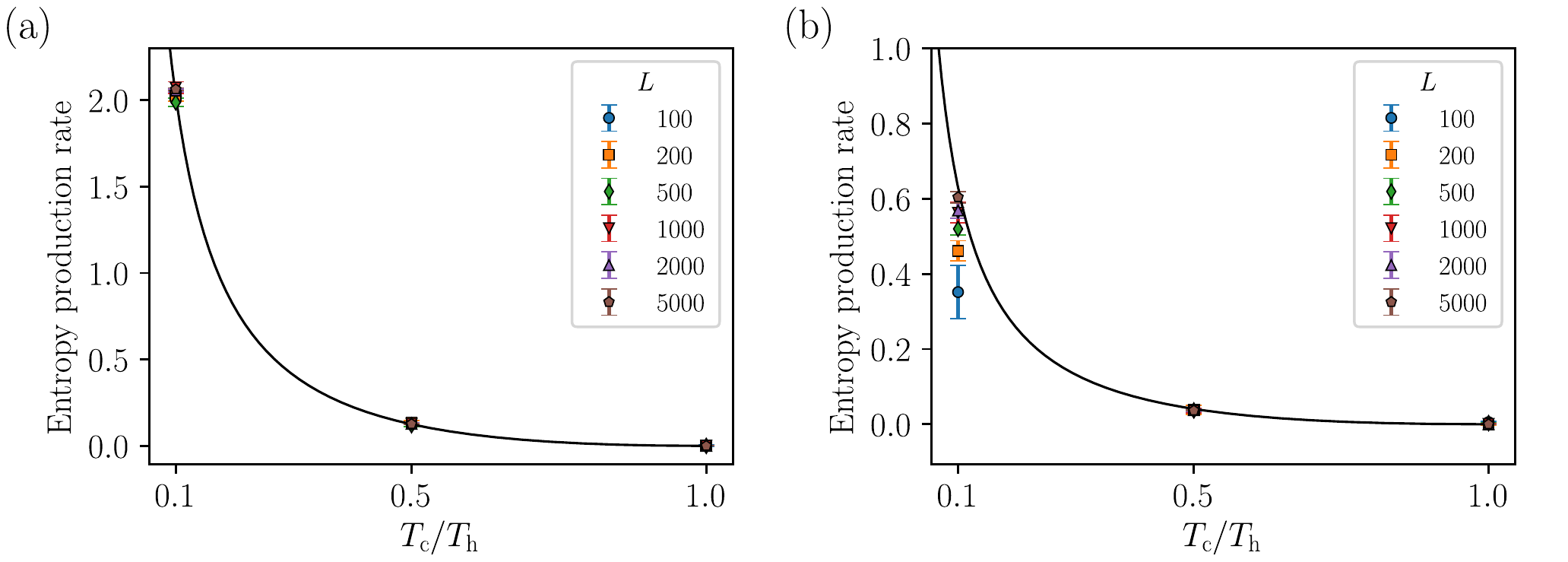}
\vskip -0.1in
\caption{Test set prediction of NEEP with $L=100,$ $200,$ $500,$ $1000,$ $2000,$ and $5000$ for the (a) two-bead and (b) five-bead models. The solid black line denotes the analytic EP rate. Error bars represent the standard deviation of estimations from five independently trained estimators.}
\vskip -0.1in
\label{figS:datasize}
\end{figure}

Next, we train NEEP for various $L$ with two- and five-bead models at $T_{\rm c}/T_{\rm h}=0.1$, $0.5$, and $1$. Figure~\ref{figS:datasize} shows that $\dot{\sigma}_{\theta^{\star}}|_{\rm test}$ can estimate $\dot{\sigma}$ with small error even when the number of training data points is small, while estimation of the five-bead model at $T_{\rm c}/T_{\rm h}=0.1$ approaches $\dot{\sigma}$ with increasing $L$.

\section{Training details}
\subsection{Bead-spring model}
\label{sec:experiment-bead}
For $N=2$ and $N=5$ bead-spring models, we report the EP rate estimation and stochastic EP using the parameter $\theta$ at the last training iteration in Fig.~\ref{fig:bead-spring}(a--e). The results of EP estimation with $\theta^{\star}$ are almost the same as the results with $\theta$ because there is no overfitting issue. The number of training iterations is $10^5$, and we evaluate $J_{\rm test}$ every $10^3$ iterations.

For the high-dimensional bead-spring model, we report $\dot{\sigma}_{\theta^{\star}}|_{\rm test}$ in Fig.~\ref{fig:bead-spring}(f). Due to the variance of each bead's position being inhomogeneous, we normalize each position in a data preprocessing step, e.g. $\bar{x}_i=(x_i - {\rm mean}(x_i))/{\rm std}(x_i)$, where the mean and standard deviation are estimated from $\mathcal{D}_{\rm train}$. The number of training iterations is $10^6$, the weight decay is $10^{-5}$, and we evaluate $J_{\rm test}$ every $10^4$ iterations. At $T_{\rm h}=10$, the respective $T_{\rm c}$ values for $N = 8$, $16$, $32$, $64$, and $128$ are $0.416997$, $0.20768$, $0.10358$, $0.05171$, and $0.02583$, where the EP rate is one.

\subsection{Discrete flashing ratchet}
\label{sec:experiment-dfr}
\begin{figure}[!ht]
\centering
\vskip 0.2in
\includegraphics[width=0.95\textwidth]{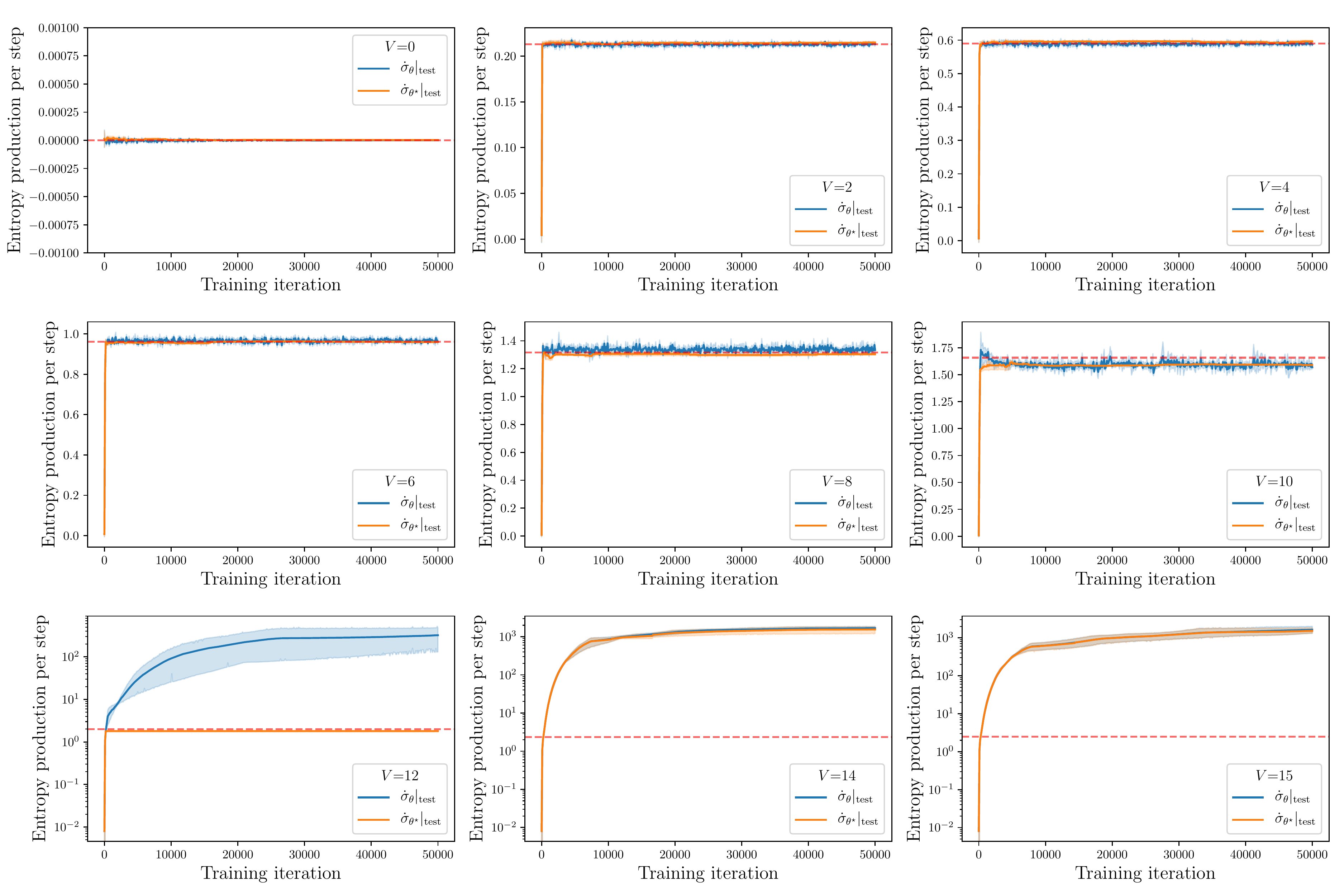}
\vskip -0.1in
\caption{Estimations of EP per step over training iteration with respect to nine different potential $V$. Red dashed lines represent the analytic EP per step $\dot{\sigma}$. The shaded areas represent the standard deviation of estimations from five independently trained estimators.}
\label{figS:ratchet}
\vskip -0.2in
\end{figure}

For the discrete flashing ratchet model with full information, we report $\dot{\sigma}_{\theta^{\star}}|_{\rm test}$ in Fig.~\ref{fig:dfr}(c). The number of training iterations is $5 \times 10^4$, and we evaluate $J_{\rm test}$ every 100 iterations. Note that these runs use a single trajectory, i.e. $M=1$ in Algorithm~\ref{alg:training}. As can be seen in Fig.~\ref{figS:ratchet}, $\dot{\sigma}_{\theta^{\star}}|_{\rm test}$ and $\dot{\sigma}_{\theta}|_{\rm test}$ have no difference when $V$ is less than 8 and approach the true value (red dashed line). At $V=12$, the overfitting issue is clearly shown as $\dot{\sigma}_{\theta}|_{\rm test}$ diverges, but  $\dot{\sigma}_{\theta^{\star}}|_{\rm test}$ converges. $\dot{\sigma}_{\theta^{\star}}|_{\rm test}$ also diverges for $V \ge 14$ because there is no $0 \to 2$ transition in the trajectory.

\begin{figure}[!ht]
\centering
\vskip 0.2in
\includegraphics[width=0.5\textwidth]{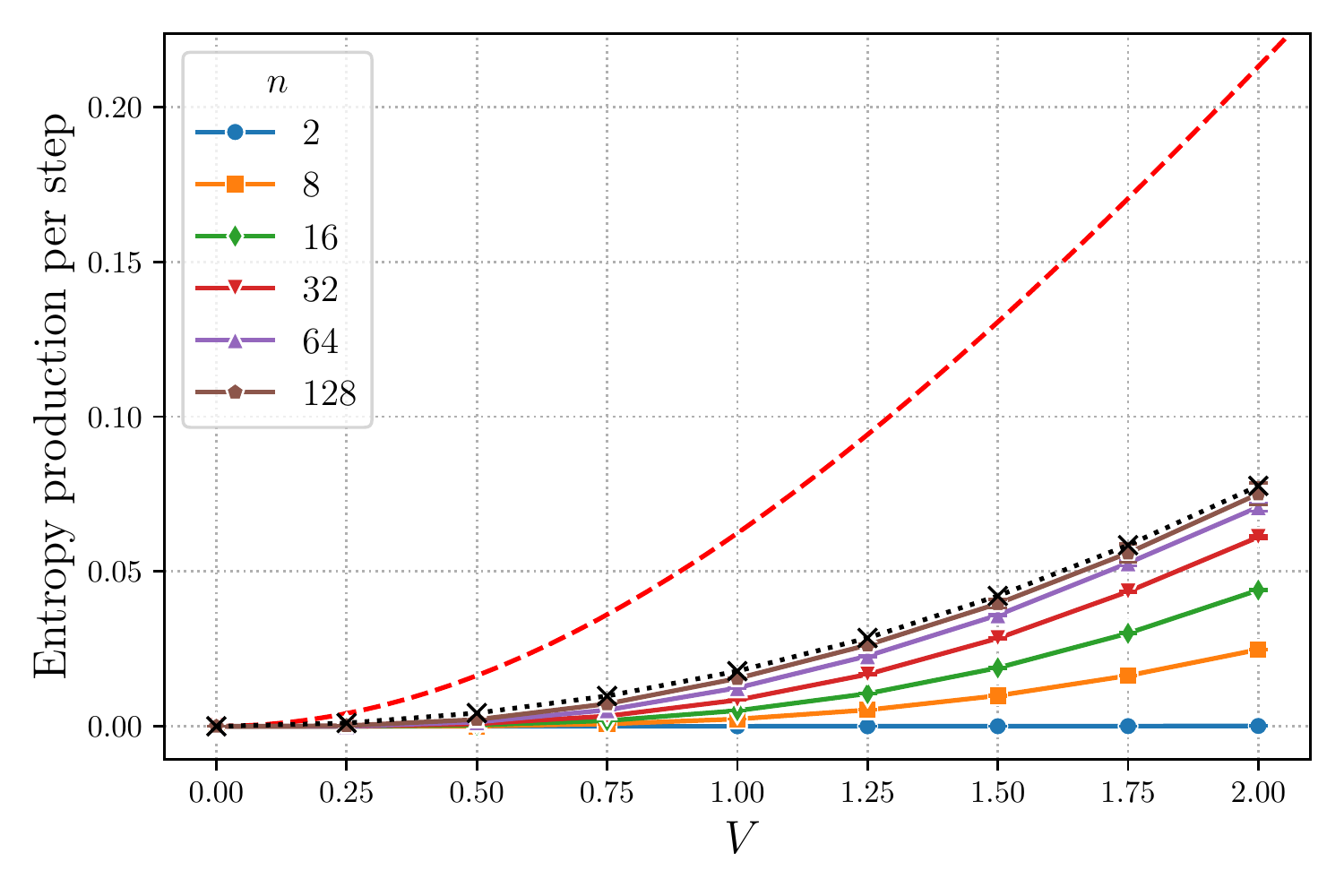}
\vskip -0.1in
\caption{Estimated coarse-grained EP per step by RNEEP ($\dot{\Sigma}_{\theta^{\star}}^n|_{\rm test}$) as a function of potential $V$ for each sequence length $n$ plotted with six different markers as shown in the legend. The black x's are the semianalytical values of $\dot{\Sigma}^{\infty}$, and the red dashed line denotes the analytic value of the actual EP per step $\dot{\sigma}$. The error bars represent the standard deviation of $\dot{\Sigma}_{\theta^{\star}}^n|_{\rm test}$ from five independently trained estimators.}
\label{figS:rneep-linear}
\vskip -0.2in
\end{figure}

\begin{figure}[!ht]
\centering
\vskip 0.2in
\includegraphics[width=\textwidth]{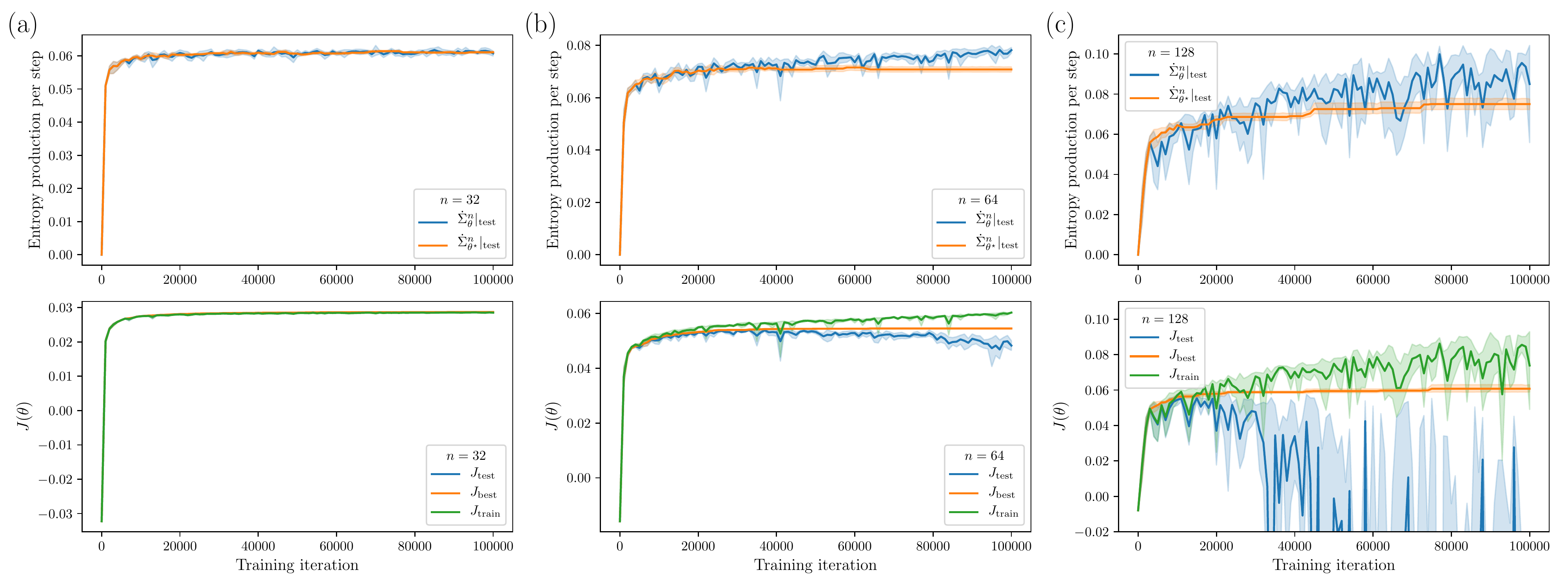}
\vskip -0.1in
\caption{Estimations of the coarse-grained EP per step $\dot{\Sigma}_{\theta}^n$ (top) and $J(\theta)$ (bottom) over training iteration by the RNEEP with (a) $n=32$, (b) $n=64$, and (c) $n=128$. The shaded areas represent the standard deviation of $\dot{\Sigma}_{\theta}^n$ and $J$ values from five independently trained estimators.}
\label{figS:rneep}
\vskip -0.2in
\end{figure}

For the discrete flashing ratchet model with partial information, we report $\dot{\Sigma}_{\theta^{\star}}^n|_{\rm test}$ in Fig.~\ref{fig:dfr-partial}(b). See Fig.~\ref{figS:rneep-linear} for a comparison between $\dot{\Sigma}_{\theta^{\star}}^n|_{\rm test}$ and the actual EP per step $\dot{\sigma}$ in linear scale. We set the dimension of the embedding vector and the number of hidden units ($H$) in the GRU to 128. See Table~\ref{table:RNEEP} for a detailed configuration of the RNEEP. We train the RNEEP with a trajectory length of $L=5\times10^7$, and the states $0', 1',$ and $2'$ are converted to $0, 1,$ and $2$ to remove the ON/OFF information. The number of training iterations is $10^5$, and we evaluate $J_{\rm test}$ every $10^3$ iterations. Figure~\ref{figS:rneep} shows the training process of RNEEP with sequence lengths $n=32$, $64$, and $128$ at potential $V=2$. For $n=32$, there is no overfitting issue: $J_{\rm train}$, $J_{\rm test}$, and $J_{\rm best}$ ($J_{\rm test}(\theta^{\star})$) are almost identical, as can be seen in Fig.~\ref{figS:rneep}(a). The overfitting issue occurs for $n=64$ (see Fig.~\ref{figS:rneep}(b)). For $n=128$, Fig.~\ref{figS:rneep}(c) shows that $J_{\rm train}$ and $J_{\rm test}$ values highly fluctuate, while $J_{\rm best}$ is robust over the training iterations; the variance of $\dot{\Sigma}_{\theta^{\star}}^n|_{\rm test}$ is also lower than $\dot{\Sigma}_{\theta}^n|_{\rm test}$.

\begin{figure}[!ht]
\centering
\vskip 0.2in
\includegraphics[width=0.5\textwidth]{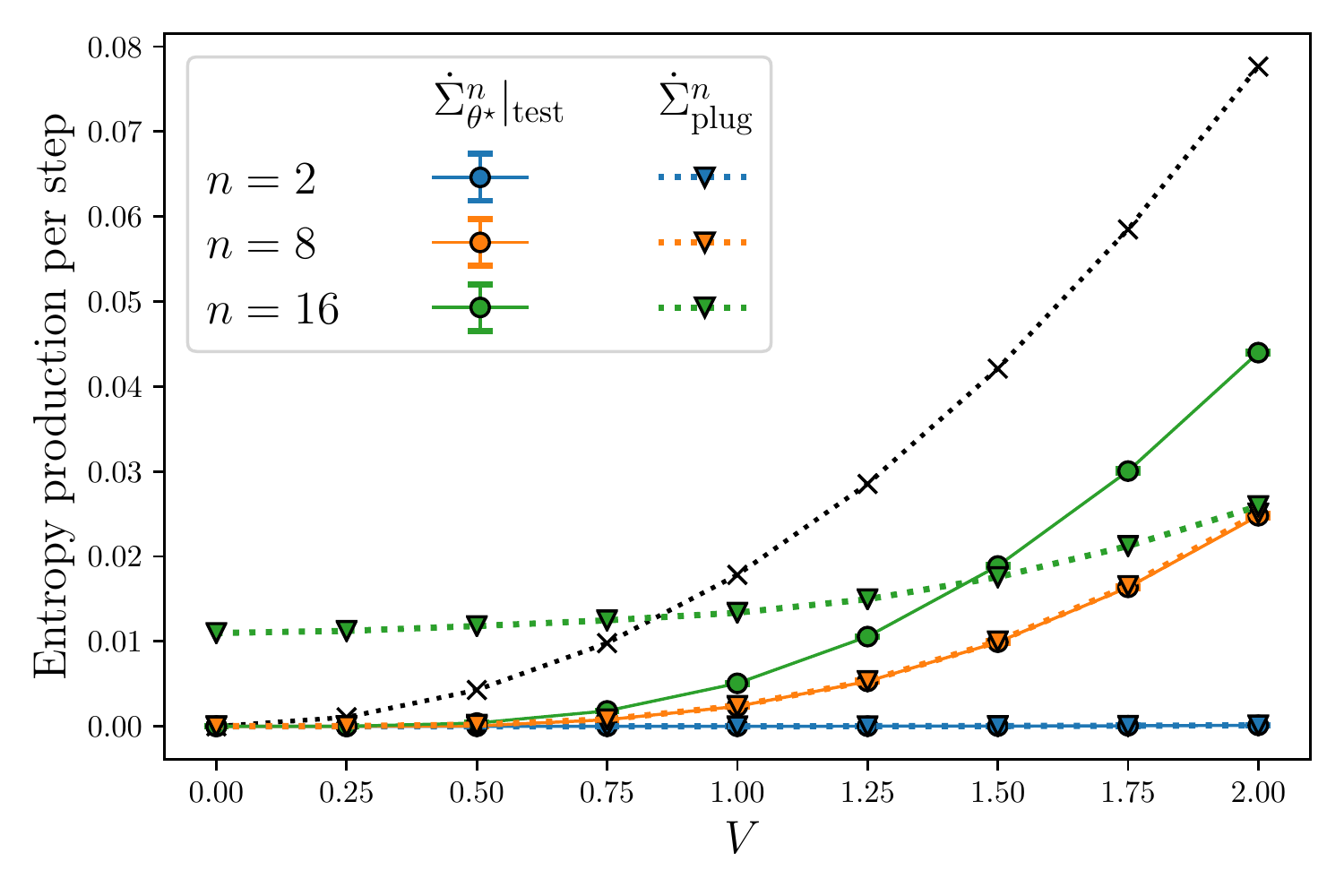}
\vskip -0.1in
\caption{Estimation of the coarse-grained EP per step by the plug-in method $\dot{\Sigma}_{\rm plug}^n$ (triangles) and RNEEP $\dot{\Sigma}_{\theta^{\star}}^n|_{\rm test}$ (circles) as a function of potential $V$ for $n=2$ (blue), $8$ (orange), and $16$ (green) in the partial information problem. The black x's are the semianalytical values of $\dot{\Sigma}^{\infty}$. The error bars represent the standard deviation of $\dot{\Sigma}_{\theta^{\star}}^n|_{\rm test}$ from five independently trained estimators.}
\label{figS:compareRNEEP}
\vskip -0.2in
\end{figure}

We now compare the RNEEP ($\dot{\Sigma}_{\theta^{\star}}^n|_{\rm test}$) with a plug-in estimator ($\dot{\Sigma}_{\rm plug}^n$) that directly estimates Eq.~\eqref{eqs:solpartialJ} by counting the frequency of $\bm{x}^n$~\cite{roldan2012entropy}. Figure~\ref{figS:compareRNEEP} shows that $\dot{\Sigma}_{\rm plug}^n$ agrees with $\dot{\Sigma}_{\theta^{\star}}^n|_{\rm test}$ for $n=2$ and $n=8$; however, at $n=16$, $\dot{\Sigma}_{\rm plug}^n$ cannot estimate well. We note that the plug-in estimator uses both training and test set data for a fair comparison with RNEEP.

\begin{table}[H]
\centering
\begin{tabular}{l|l|l}
\hline
\multicolumn{3}{c}{
\centering
\textbf{RNEEP with sequence length $n$}} \\

\hline
  Layer & \makecell{Output dim} & Activation function\\
\hline
Input $x$ & $\{0, 1, 2\}$ &  \\
Embedding & $H$ &  \\
Concatenate $\bm{x}^n$ & $ n \times H$ &  \\
GRU($\bm{x}^n$) & $ n \times H $ & None \\
Average & $H$ & None \\
Output layer & $1$ &  None\\
\hline
\end{tabular}
\caption{RNEEP configuration for the partial information problem.}
\label{table:RNEEP}
\end{table}

\newpage
\section{Proof for coarse-grained entropy production}
In this section, we prove that our objective function $J(\theta)$ has a maximum value when $\Delta S_{\theta}$ is the stochastic coarse-grained EP.
In steady-state, the objective function $J(\theta)$ can be written as
\begin{align}
    J[h] = \sum_{\bm{x}^n} \sum_{\bm{\eta}^n} 
        p(\bm{x}^n, \bm{\eta}^n) \left[ h(\bm{x}^n) - h(\widetilde{\bm{x}}^n) - e^{-(h(\bm{x}^n) - h(\widetilde{\bm{x}}^n))}
    \right],
    \label{eqs:rneep_J}
\end{align}
where $n$ is the length of sequence $\bm{x}^n$ and 
\begin{align}
    \bm{x}^n &= (x_1, x_2, ..., x_n),\;\; \widetilde{\bm{x}}^n = (x_n, ..., x_2, x_1), \nonumber \\
    \bm{\eta}^n &= (\eta_1, \eta_2, ..., \eta_n),\;\; \widetilde{\bm{\eta}}^n = (\eta_n, ..., \eta_2, \eta_1). \nonumber
\end{align}
The maximum condition for Eq.~\eqref{eqs:rneep_J} can be obtained as follows:
\begin{align}\label{eqs:partialJ}
0 =& \partial_{h(\bm{x}'^n)} J[h]\\
  =&\sum_{\bm{x}^n} \sum_{\bm{\eta}^n} 
    p(\bm{x}^n, \bm{\eta}^n) \left( \delta_{\bm{x}^n, \bm{x}'^n} - \delta_{\widetilde{\bm{x}}^n, \bm{x}'^n}\right) \left[1 + e^{-(h(\bm{x}^n)-h(\widetilde{\bm{x}}^n))} \right] 
    \nonumber\\
    =& \sum_{\bm{\eta}^n} p(\bm{x}'^n, \bm{\eta}^n) \left[1 + e^{-(h(\bm{x}'^n)-h(\widetilde{\bm{x}}'^n))}  \right] - \sum_{\bm{\eta}^n} p(\widetilde{\bm{x}}'^n, \bm{\eta}^n) \left[1 + e^{-(h(\widetilde{\bm{x}}'^n) - h(\bm{x}'^n))}  \right] \nonumber\\ 
    =& \left[1 + e^{-(h(\widetilde{\bm{x}}'^n) - h(\bm{x}'^n))} \right] \left[ e^{-(h(\bm{x}'^n)-h(\widetilde{\bm{x}}'^n))} \sum_{\bm{\eta}^n} p(\bm{x}'^n, \bm{\eta}^n) - \sum_{\bm{\eta}^n} p(\widetilde{\bm{x}}'^n, \bm{\eta}^n) \right]. \nonumber
\end{align}
Then the solution for the optimization problem is
\begin{align}\label{eqs:solpartialJ}
    h(\bm{x}'^n) - h(\widetilde{\bm{x}}'^n) = - \ln{\frac{\sum_{\widetilde{\bm{\eta}}^n} p(\widetilde{\bm{x}}'^n, \widetilde{\bm{\eta}}^n)}{\sum_{\bm{\eta}^n} p(\bm{x}'^n, \bm{\eta}^n)}} = -\ln{\frac{p(\widetilde{\bm{x}}'^n)}{p(\bm{x}'^n)}}.
\end{align}